\documentstyle[11pt,a4]{article}
\begin{document}
\title{Cosmology with bulk pressure }
\author{ Winfried Zimdahl$^{1}$, Diego Pav\'{o}n$^{2}$  and
Josep Triginer$^{2}$
\\
$^{1}$Fakult\"at f\"ur Physik, Universit\"at Konstanz,
PF 5560 M678\\
D-78434  Konstanz, Germany\\
and\\
$^{2}$Departament de F\'{\i}sica\\
Universitat Aut\`{o}noma de Barcelona\\
08193 Bellaterra (Barcelona), Spain}
\date{\today}
\maketitle
\abstract{
It is well known that cosmological particle
production processes may phenomenologically be
described in terms of effective viscous pressures.
We investigate the consistency of this approach
on the level of relativistic kinetic theory,
using a Boltzmann equation with an additional source
term  that leads to a change in the number of gas particles.
For a simple creation rate model we find the
conditions for collisional equilibrium of a
Maxwell-Boltzmann gas with increasing particle
number and discuss their cosmological implications.
We also comment on the possibility of a bulk pressure
driven inflationary phase within causal, nonequilibrium
thermodynamics.
}
\section{Introduction}
Entropy producing processes have played an important role
in the evolution of the early universe.
On the level of fluid cosmology, the simplest
phenomenon connected with a nonvanishing entropy production is a  
bulk viscosity.
A bulk viscous pressure is the only dissipative effect
that is compatible with the symmetry requirements of the
homogeneous and isotropic Friedmann-Lema\^{\i}tre-Robertson-Walker  
models. For the possible physical origin of bulk pressures in the  
expanding universe see \cite{ZMN} and references therein.\\
On the other hand,
there exist attempts to use effective viscous pressures in order
to model particle production proceses
in the early universe
(see  references in \cite{TZP}). For certain quantum processes in  
the early universe `viscosity functions' have been calculated,  
opening the possibility of an effective imperfect fluid description  
of these phenomena.\\
Generally, any production process will give rise to source terms in  
the macroscopic fluid balances.
The interpretation of these contributions
as effective bulk pressures uses the fact that a source term in the  
energy balance of a fluid may formally be rewritten in terms of an  
effective bulk pressure.
The advantage of a rewriting like this is obvious:
An energy-momentum balance with a nonzero source term violates the  
integrability conditions of Einstein's equations, while the  
energy-momentum tensor of an imperfect fluid is a reasonable  
right-hand side of the field equations.
Moreover, this approach allows us to calculate the backreaction of  
the corresponding process on the cosmological dynamics.\\
We shall investigate here whether the approach of regarding  
particle production processes as equivalent to viscous pressures is   
consistent with a kinetic description.
In section 2 we sketch the general kinetic theory of a simple gas  
with nonconserved particle number and introduce an effective rate  
approximation for the production term. The conditions for  
collisional equilibrium are found in section 3, while the above  
mentioned effective viscous pressure approach is established in  
section 4.
The backreaction of particle production processes on the  
cosmological dynamics under the conditions of collisional  
equilibrium is considered in section 5.
In section 6 we drop the equilibrium conditions and comment on the  
possibility  of a bulk pressure driven inflationary phase within   
causal, nonequilibrium thermodynamics.
\section{Basic kinetic theory}
The one-particle distribution function $ f = f\left(x,p\right)$ 
of a relativistic gas with varying particle number 
is supposed to obey the equation \cite{TZP}
\begin{equation}
L\left[f\right] \equiv 
p^{i}f,_{i} - \Gamma^{i}_{kl}p^{k}p^{l}
\frac{\partial f}{\partial
p^{i}}
 = C\left[f\right] + S\left(x, p\right)\ .\label{1}
\end{equation}
$L\left[f\right] $ is the Liouville operator and
$C[f]$ is Boltzmann's collision term.
The source term $S$ takes into account the fact that the  
distribution function $f$ may additionally vary due to creation or  
decay processes, supposedly of quantum origin.
On the level of classical kinetic theory we regard this term as a given
input quantity.
Assuming a linear coupling of the source term $S$ to $f$ and  
requiring a macroscopic description of the production process in  
terms of the zeroth, first and second moments of $f$ only, provides  
us with the expression
\begin{equation}
S\left(x, p\right) =  \left(- \frac{u_{a}p^{a}}{\tau\left(x\right)}  
+ \nu\left(x\right)\right)
f\left(x, p\right) ,
\label{2}
\end{equation}
for the source term in (\ref{1}),
where $u ^{a}$ is the macroscopic 4-velocity.
The creation process is characterized by the spacetime
functions $\tau $ and $\nu$.
\section{Collisional equilibrium}
As in the case $S = 0$, the condition for collisional equilibrium
implies the structure
\begin{equation}
f\left(x, p\right) = f^{0}\left(x, p\right) =
\exp{\left[\alpha + \beta_{a}p^{a}\right] }
\mbox{ , }\label{3}
\end{equation}
of the distribution function. Using the latter expression together  
with (\ref{2}) in  (\ref{1}) yields the equilibrium conditions
\begin{equation}
\dot{\alpha}=\frac{1}{\tau}\ ,\ \ \ h^{a}_{b}\alpha_{,a}=0,\ \ \
\beta_{(a;b)}= - \frac{\nu}{m ^{2}}g_{ab}\ , \
\label{4}
\end{equation}
which are less restrictive than in the case of particle number  
conservation.
Different from the case without particle production ($\tau ^{-1} =  
\nu = 0$)
we find that $\alpha$ may change along the fluid flow lines and that 
$\beta_{a} \equiv u _{a}/T$ obeys  a conformal Killing equation for  
 $m \neq 0$.
The latter property implies the temperature behaviour
$\dot{T}/T = - \Theta /3 $
that coincides with the temperature law for massless particles  
(radiation) ($\Theta = u^{a}_{;a}$ is the fluid expansion).
We conclude that
{\it radiation and matter in the expanding Universe may be in equilibrium
at the same temperature, provided the particle number of the matter
component is allowed to change at a specific rate}.
It is well known, that for conserved particle numbers  an  
equilibrium between both components is
impossible.\\
A further consequence of the equilibrium conditions (\ref{4}) is a  
modification of Tolman's relation \cite{TZP}. For nonvanishing  
particle production {\it the quantity $T\sqrt{-g _{00}}$ is allowed  
to be time dependent}. However, like $\alpha$ in (\ref{4}), it has  
to be spatially constant.
\section{The effective viscous pressure approach}
Let, as usual, the energy-momentum tensor $T ^{ik}$ be defined as  
the second moment of the distribution function (\ref{3}).
{\it The effective viscous pressure approach then amounts to the  
replacement of the nonconserved energy-momentum tensor
$T^{ik}=\rho u^{i}u^{k}+p h^{ik}$ with
$\dot{\rho} + \Theta\left(\rho + p\right) 
= \rho /\tau + \nu n  $
by the conserved energy-momentum tensor
$\hat{T}^{ik}=\rho u^{i}u^{k}+(p+\pi)h^{ik}$
with
$\dot{\rho} + \Theta\left(\rho + p + \pi \right) 
= 0 $, where
$\Theta \pi  = - \rho / \tau - \nu n$}.
We find this approach to be  consistent for homogeneous spacetimes  
but not necessarily for inhomogeneous ones \cite{TZP}.
\section{Backreaction on the cosmological dynamics}
A nonvanishing effective bulk pressure influences the entire  
cosmological dynamics.
Restricting ourselves to the homogeneous, isotropic and spatially  
flat case, the relation $2\dot{H} /H  = \dot{\rho}/\rho$ is valid,    

where $H$ is the Hubble parameter $H = \Theta /3$.
The functions $\nu$ and $\tau $ are fixed by the conditions  
(\ref{4}) and by the additional assumption of `adiabatic' particle  
production \cite{ZTP}.
With $H \equiv  \dot{a}/a$, where $a$ is the scale factor of the  
Robertson-Walker metric, we find (see \cite{ZTP}) that for massive  
particles
($m \gg T$) the scale factor behaves like
$a \propto t^{4/3}$
instead of the familiar $a \propto t^{2/3}$ for $\tau ^{-1} = \nu =  
0$.
{\it Massive particles with adiabatically changing  particle number  
are allowed to be in collisional equilibrium only in a power-law  
inflationary universe}.
Introducing a `$\gamma$-law' in the usual form $p = (\gamma - 1) \rho $, 
one obtains accelerated expansion, i.e., $\dot{a}/a > 0$, in the  
interval
$1 \leq \gamma \leq \gamma _{cr}$ with
$\gamma_{cr} \approx 1.09$.
In the opposite limit of massless particles the equilibrium  
conditions require $\tau ^{-1} = \nu = 0$, i.e., there is no  
adiabatic production of relativistic particles.\\
We also investigated whether a phase of exponential inflation is  
consistent with the conditions for collisional equilibrium  
\cite{ZTP}.
In our setting exponential inflation may be realized  by  
`nonadiabatic' particle production.
It turns out that {\it the equilibrium conditions (\ref{4}) are  
compatible with a de Sitter phase only for a time interval of the  
order of the Hubble time}.
For larger times exponential inflation violates the second law of  
thermodynamics, i.e., the de Sitter phase becomes thermodynamically  
unstable.
\section{Causal thermodynamics and inflation}
Up to now our analysis was based on the equilibrium conditions  
(\ref{4}).
Now we drop these conditions and investigate the role of bulk  
pressures within causal, nonequilibrium thermodynamics (see  
\cite{Maar,ZPRD} and references therein).
The viscous pressure becomes a dynamical degree of freedom on its  
own and the cosmological dynamics is determined by a second-order  
equation for the Hubble parameter $H$.
A basic ingredient of the theory is a nonvanishing relaxation time. 
There exists a debate on whether deviations from thermodynamical  
equilibrium may drive a de Sitter phase.
Assuming the applicability of the causal, nonequilibrium  
thermodynamics also far from equilibrium,
it may be shown that {\it the causal evolution equation for $H$  
admits stable, inflationary solutions with relaxation times of the  
order of the Hubble time,
i.e., the nonequilibrium may be `frozen in'}.
Provided, the  bulk pressure is again due to particle production  
processes, one finds that the temperature remains approximately  
constant during the de Sitter phase (no supercooling!). The latter  
feature is connected with an exponential increase of the comoving  
entropy.
{\it While the standard scenario of an inflationary universe  
comprises an adiabatic  de Sitter phase and a subsequent reheating  
period, in which all the entropy in the universe is produced, the  
entropy production takes place during  the de Sitter stage itself in  
the present picture}.


\begin{thebibliography}{99}
\bibitem{ZMN} W. Zimdahl, {\em Mon. Not. R. Astr. Soc.} {\bf 280},  
1239 (1996)
\bibitem{TZP} J. Triginer, W. Zimdahl, D. Pav\'{o}n,
{\em Class. Quantum Grav.} {\bf 13}, 403 (1996)
\bibitem{ZTP}  W. Zimdahl, J. Triginer,  D. Pav\'{o}n, submitted to 
{\em Phys. Rev. D}
\bibitem{Maar} R. Maartens, {\em Class. Quantum Grav.} {\bf 12},  
1455 (1995)
\bibitem{ZPRD} W. Zimdahl,
{\em Phys. Rev.} {\bf D 53}, 5483 (1996)
\end{thebibliography}
\end{document}